# Interval edge-colorings of cubic graphs

Petros A. Petrosyan

Institute for Informatics and Automation
Problems of NAS of RA,
Department of Informatics and Applied
Mathematics, YSU,
Yerevan, Armenia
e-mail: pet_petros@ipia.sci.am

## ABSTRACT

An edge-coloring of a multigraph $G$ with colors $1,2,\ldots,t$ is called an interval $t$-coloring if all colors are used, and the colors of edges incident to any vertex of $G$ are distinct and form an interval of integers. In this paper we prove that if $G$ is a connected cubic multigraph (a connected cubic graph) that admits an interval $t$-coloring, then $t \leq |V(G)|+1$ ($t \leq |V(G)|$), where $V(G)$ is the set of vertices of $G$. Moreover, if $G$ is a connected cubic graph, $G \neq K_4$, and $G$ has an interval $t$-coloring, then $t \leq |V(G)|-1$. We also show that these upper bounds are sharp. Finally, we prove that if $G$ is a bipartite subcubic multigraph, then $G$ has an interval edge-coloring with no more than four colors.

## 1. INTRODUCTION

In this paper we consider graphs which are finite, undirected, and have no loops or multiple edges and multigraphs which may contain multiple edges but no loops. Let $V(G)$ and $E(G)$ denote the sets of vertices and edges of a graph $G$, respectively. An $(a,b)$-biregular bipartite graph $G$ is a bipartite graph $G$ with the vertices in one part all having degree $a$ and the vertices in the other part all having degree $b$. The degree of a vertex $v \in V(G)$ is denoted by $d_G(v)$, the maximum degree of a vertex in $G$ by $\Delta(G)$ and the chromatic index of $G$ by $\chi'(G)$. The terms and concepts that we do not define can be found in [7,10].

An interval $t$-coloring of a multigraph $G$ is an edge-coloring of $G$ with colors $1,2,\ldots,t$ such that at least one edge of $G$ is colored by color $i$, $i=1,2,\ldots,t$, and the edges incident to each vertex $v \in V(G)$ are colored by $d_G(v)$ consecutive colors. A multigraph $G$ is interval-colorable if there is $t \geq 1$ for which $G$ has an interval $t$-coloring. The set of all interval-colorable multigraphs is denoted by $\mathfrak{N}$. For a multigraph $G \in \mathfrak{N}$, the least and the greatest values of $t$ for which $G$ has an interval $t$-coloring are denoted by $w(G)$ and $W(G)$, respectively.

The concept of interval edge-coloring of multigraphs was introduced by Asratian and Kamalian [4]. In [4,5], they proved the following two theorems.

**Theorem 1.** If $G$ is a regular graph, then $G \in \mathfrak{N}$ if and only if $\chi'(G) = \Delta(G)$.

**Theorem 2.** If $G$ is a connected triangle-free graph and $G \in \mathfrak{N}$, then $W(G) \leq |V(G)|-1$.

**Corollary 1.** If $G$ is a connected bipartite graph and $G \in \mathfrak{N}$, then $W(G) \leq |V(G)|-1$.

Note that this upper bound is tight for complete bipartite graphs $K_{m,n}$, since $K_{m,n} \in \mathfrak{N}$ and $W(K_{m,n}) = m+n-1$ [11]. Nevertheless, for some bipartite graphs this upper bound can be improved. Recently, Asratian and Casselgren [3] proved the following

**Theorem 3.** If $G$ is a connected $(a,b)$-biregular bipartite graph with $|V(G)| \geq 2(a+b)$ and $G \in \mathfrak{N}$, then
$$W(G) \leq |V(G)|-3.$$

For general graphs, Kamalian proved the following

**Theorem 4.** [12] If $G$ is a connected graph and $G \in \mathfrak{N}$, then
$$W(G) \leq 2|V(G)|-3.$$

Note that the upper bound is tight for $K_2$, but if $G \neq K_2$, then this upper bound can be improved.

**Theorem 5.** [8] If $G$ is a connected graph with $|V(G)| \geq 3$ and $G \in \mathfrak{N}$, then
$$W(G) \leq 2|V(G)|-4.$$

For regular graphs, Kamalian and Petrosyan proved the following

**Theorem 6.** [13] If $G$ is a connected $r$-regular graph with $|V(G)| \geq 2r+2$ and $G \in \mathfrak{N}$, then
$$W(G) \leq 2|V(G)|-5.$$

On the other hand, in [16], Petrosyan proved the following theorem.

**Theorem 7.** For any $\varepsilon > 0$, there is a connected graph $G$ such that $G \in \mathfrak{N}$ and $W(G) \geq (2-\varepsilon)|V(G)|$.

For planar graphs, the coefficient in upper bounds of Theorems 4-6 was improved by Axenovich. In [6], she proved the following

**Theorem 8.** If $G$ is a connected planar graph and $G \in \mathfrak{N}$, then
$$W(G) \leq \frac{11}{6}|V(G)|.$$

In this paper we investigate interval edge-colorings of cubic graphs and multigraphs. We also consider interval edge-colorings of bipartite subcubic multigraphs.

## 2. MAIN RESULTS

First, we give an upper bound on $W(G)$ for interval-colorable connected cubic multigraphs $G$.

**Theorem 9.** If $G$ is a connected cubic multigraph and $G \in \mathfrak{N}$, then
$$W(G) \leq |V(G)| + 1.$$

Note that the upper bound in Theorem 9 is tight. The following theorem holds.

**Theorem 10.** For any $n \geq 2$, there exists a connected cubic multigraph $G$ with $|V(G)| = 2n$ such that $G \in \mathfrak{N}$ and
$$W(G) = |V(G)| + 1.$$

Next, we show that if $G$ is a connected cubic graph and $G \in \mathfrak{N}$, then $W(G) \leq |V(G)|$.

**Theorem 11.** If $G$ is a connected cubic graph and $G \in \mathfrak{N}$, then $W(G) \leq |V(G)|$. Moreover, if $G \neq K_4$, then
$$W(G) \leq |V(G)| - 1.$$

Note that the upper bound in Theorem 11 is tight, too. The following theorem holds.

**Theorem 12.** For any $n \geq 3$, there exists a connected cubic graph $G$ with $|V(G)| = 2n$ such that $G \in \mathfrak{N}$ and
$$W(G) = |V(G)| - 1.$$

It is well-known that the four color theorem [1,2] is equivalent to the statement that every bridgeless cubic planar graph is $3$-edge-colorable [10]. From here and taking into account Theorem 1, we obtain the following

**Theorem 13.** If $G$ is a bridgeless planar cubic graph, then $G \in \mathfrak{N}$ and $w(G) = 3$.

$K_4$ is an example of bridgeless planar cubic graph with $W(K_4) = |V(K_4)| = 4$, but if $G$ is a 2-connected planar cubic graph and $G \neq K_4$, then, by Theorem 11, we have $W(G) \leq |V(G)| - 1$. Moreover, we show that the following theorem holds.

**Theorem 14.** For any $n \geq 3$, there exists a 2-connected planar cubic graph $G$ with $|V(G)| = 2n$ such that
$$W(G) = |V(G)| - 1.$$

We also consider cubic Halin graphs. A Halin graph is a planar graph constructed from a plane embedding of a tree with at least four vertices and with no vertices of degree 2, by connecting all the leaves of the tree with a cycle, traversing the leaves in the order given by planar embedding of the tree. In particular, we proved the following result.

**Theorem 15.** If $G$ is a cubic Halin graph, then $G \in \mathfrak{N}$ and $w(G) = 3$. Moreover, for any $n \geq 2$, there exists a cubic Halin graph $G$ with $|V(G)| = 2n$ such that
$$W(G) \geq \frac{|V(G)|}{2} + 2.$$

However, there are interval-colorable cubic graphs $G$ for which the value of the parameter $W(G)$ is close to $\frac{|V(G)|}{2}$. For example, in [15], it was proved that for Möbius ladders $M_{2n}$ ($n \geq 2$) with $2n$ vertices the following theorem holds.

**Theorem 16.** For any $n \geq 2$,
1. $M_{2n} \in \mathfrak{N}$,
2. $w(M_{2n}) = 3$,
3. $W(M_{2n}) = n + 2$,
4. if $w(M_{2n}) \leq t \leq W(M_{2n})$, then the cubic graph $M_{2n}$ has an interval $t$-coloring.

In [14], a similar result was proved for $n$-prism graphs $C_n \square K_2$ ($n \geq 3$) with $2n$ vertices. In particular, Khchoyan [14] proved the following

**Theorem 17.** For any $n \geq 2$,
1. $C_n \square K_2 \in \mathfrak{N}$,
2. $w(C_n \square K_2) = 3$,
3. $W(C_n \square K_2) = n + 2$,
4. if $w(C_n \square K_2) \leq t \leq W(C_n \square K_2)$, then the cubic graph $C_n \square K_2$ has an interval $t$-coloring.

In the same work [14], the author considered a ring of diamonds $D_k$ (a ring with $k$ diamonds and $k \geq 2$). The graph $K_4 - e$ is called a diamond, and a ring of $k$ diamonds is a sequence of $k$ diamonds in which consecutive diamonds are connected. Clearly, a ring of diamonds $D_k$ is a 2-

connected planar cubic graph. For these graphs, Khchoyan proved the following

**Theorem 18.** For any $k \geq 2$,

1. $D_k \in \mathfrak{N}$,
2. $w(D_k) = 3$,
3. $W(D_k) \geq \frac{3}{2}k + 4$, if $k$ is even,
   $W(D_k) \geq 3\left\lceil \frac{k}{2} \right\rceil + 2$, if $k$ is odd,
4. if $k$ is even and $w(D_k) \leq t \leq \frac{3}{2}k + 4$, then $D_k$ has an interval $t$−coloring,
   if $k$ is odd and $w(D_k) \leq t \leq 3\left\lceil \frac{k}{2} \right\rceil + 2$, then $D_k$ has an interval $t$−coloring.

Finally, we consider bipartite multigraphs. In [9], Hansen proved the following theorem for bipartite subcubic graphs.

**Theorem 19.** If $G$ is a bipartite graph with $\Delta(G) \leq 3$, then $G \in \mathfrak{N}$ and $w(G) \leq 4$.

We show that this theorem also holds for bipartite subcubic multigraphs. More precisely, the following theorem holds.

**Theorem 20.** If $G$ is a bipartite multigraph with $\Delta(G) \leq 3$, then $G \in \mathfrak{N}$ and $w(G) \leq 4$.

By Theorems 9 and 20, we obtain the following

**Corollary 2.** If $G$ is a connected bipartite cubic multigraph, then $G \in \mathfrak{N}$ and $W(G) \leq |V(G)| + 1$. Moreover, for any $n \geq 2$, there exists a connected bipartite cubic multigraph $G$ with $|V(G)| = 2n$ and $W(G) = |V(G)| + 1$.

On the other hand, by Theorems 3 and 19, we obtain the following

**Corollary 3.** If $G$ is a connected bipartite cubic graph with $|V(G)| \geq 12$, then $G \in \mathfrak{N}$ and
$$W(G) \leq |V(G)| - 3.$$

## ACKNOWLEDGEMENT

We would like to express our gratitude to H. Tananyan and V. Mkrtchyan for useful discussions on the subject.

## REFERENCES

[1] K. Appel, W. Haken, "Every planar map is four colorable, Part I, Discharging", *Illinois J. Math. 21,* pp. 429-490, 1977.

[2] K. Appel, W. Haken, J. Koch, "Every planar map is four colorable, Part II, Reducibility", *Illinois J. Math. 21*, pp. 491-567, 1977.

[3] A.S. Asratian, C.J. Casselgren, "On interval edge colorings of $(a,b)$−biregular bipartite graphs", *Discrete Math. 307*, pp. 1951-1956, 2006.

[4] A.S. Asratian, R.R. Kamalian, "Interval colorings of edges of a multigraph", *Appl. Math. 5*, pp. 25-34, 1987.

[5] A.S. Asratian, R.R. Kamalian, "Investigation on interval edge-colorings of graphs", *J. Combin. Theory Ser. B 62*, pp. 34-43, 1994.

[6] M.A. Axenovich, "On interval colorings of planar graphs", *Congr. Numer. 159*, pp. 77-94, 2002.

[7] J.A. Bondy, U.S.R. Murty, "Graph Theory", Springer, 2008.

[8] K. Giaro, M. Kubale, M. Malafiejski, "Consecutive colorings of the edges of general graphs", *Discrete Math. 236*, pp. 131-143, 2001.

[9] H.M. Hansen, "Scheduling with minimum waiting periods", Master's Thesis, Odense University, 1992.

[10] F. Harary, "Graph Theory", Addison-Wesley, Reading MA, 1969.

[11] R.R. Kamalian, "Interval colorings of complete bipartite graphs and trees", *Preprint of the Comp. Cen. of Acad. Sci. of Armenian SSR*, 1989.

[12] R.R. Kamalian, "Interval edge-colorings of graphs", Doctoral Thesis, Novosibirsk, 1990.

[13] R.R. Kamalian, P.A. Petrosyan, "A note on interval edge-colorings of graphs", *Graphs and Combinatorics*, 2011, under review.

[14] A. Khchoyan, "Interval edge-colorings of subcubic graphs and multigraphs", Yerevan State University, BS thesis, 30p., 2010.

[15] P.A. Petrosyan, "Interval edge colorings of Möbius ladders", *Proceedings of the CSIT Conference*, pp. 146-149, 2005.

[16] P.A. Petrosyan, "Interval edge-colorings of complete graphs and $n$−dimensional cubes, *Discrete Math. 310*, pp. 1580-1587, 2010.